\documentclass[twocolumn]{aastex6}

\newcommand{\Teff}{$T\mathrm{\hspace*{-0.4ex}_{eff}}$\,}
\newcommand{\logg}{$\log\,g$\hspace*{0.5ex}}

\def\re{RE\,0503$-$289}

\begin{document}


\title{Detection of forbidden line components of lithium-like carbon in stellar spectra}


\author{Klaus Werner$^1$, Thomas Rauch$^1$, Denny Hoyer$^1$, and Pascal Quinet$^{2,3}$}
\affil{$^1$Institute for Astronomy and Astrophysics, Kepler Center for Astro and
Particle Physics,  Eberhard Karls University T\"ubingen, Sand~1, 72076
T\"ubingen, Germany
\\$^2$Physique Atomique et Astrophysique, Universit\'e de Mons -- UMONS, 7000 Mons, Belgium
\\$^3$IPNAS, Universit\'e de Li\`ege, Sart Tilman, 4000 Li\`ege, Belgium}

\begin{abstract}

We report the first identification of forbidden line
  components from an element heavier than helium in the spectrum of
  astrophysical plasmas. As yet, these components were identified only
  in laboratory plasmas and not in astrophysical objects. Forbidden
  components are well known for neutral helium lines in hot stars,
  particularly in helium-rich post-AGB stars and white dwarfs. We
  discovered that two hitherto unidentified lines in the ultraviolet
  spectra of hot hydrogen-deficient (pre-) white dwarfs can be
  identified as forbidden line components of triply ionized carbon
  (\ion{C}{4}). The forbidden components (3p--4f and 3d--4d) appear
  in the blue and red wings of the strong, Stark broadened 3p--4d and
  3d--4f lines at 1108\,\AA\ and 1169\,\AA, respectively. They are
  visible over a wide effective temperature range
  ($60\,000-200\,000$\,K) in helium-rich (DO) white dwarfs and PG\,1159
  stars that have strongly oversolar carbon abundances.

\end{abstract}


\section{Introduction}
\label{intro}

Forbidden line components are atomic transitions with $\Delta \ell
\not= \pm 1$, where $\ell$ is the angular quantum number. They are
associated with the mixing of upper states induced by the plasma
electric microfield, leading to transitions that are normally
disallowed by the selection rules for electric dipole transitions.
This effect should not be confused with forbidden lines associated
with magnetic dipole, electric quadrupole or other higher multipole
transitions which are well known tools for analysing emission lines
from thin astrophysical plasmas, e.g., a multitude of forbidden lines
in planetary nebulae \citep{1927Natur.120..473B} and He-like triplets
in X-ray spectra of stellar coronae \citep{1969Natur.221..947G}. The
forbidden line components investigated here are not restricted to low
densities because they do not involve metastable states. In contrast,
they appear as absorption lines at high densities when line broadening
by the Stark effect is important.

\begin{figure*}[bth]
 \centering\includegraphics[height=0.81\textheight]{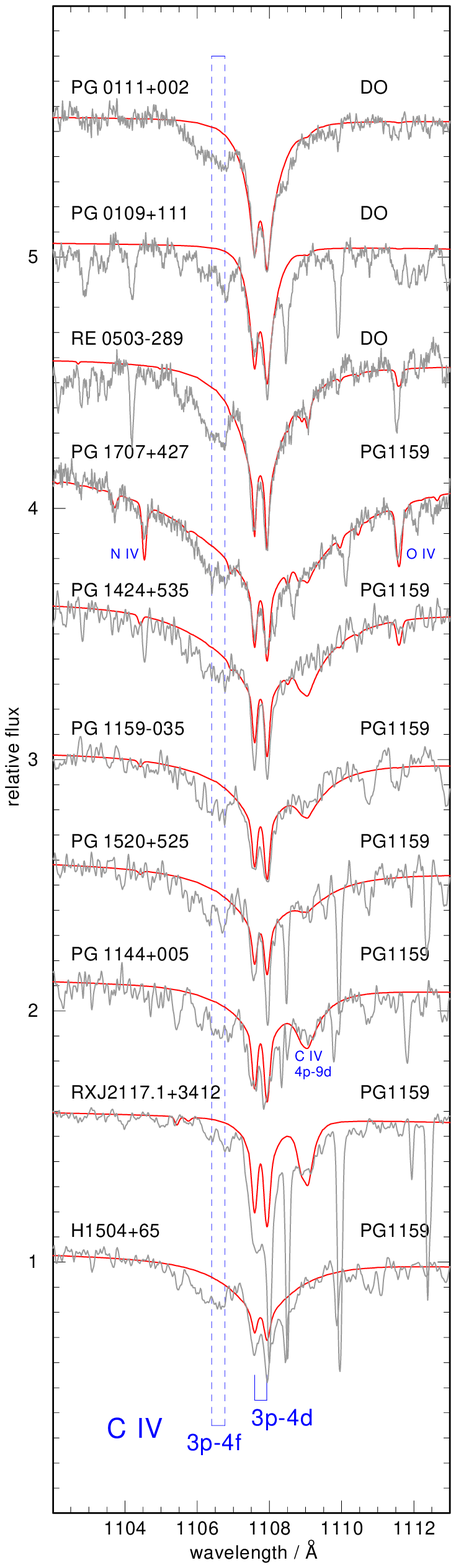}\hspace{1cm}
           \includegraphics[height=0.81\textheight]{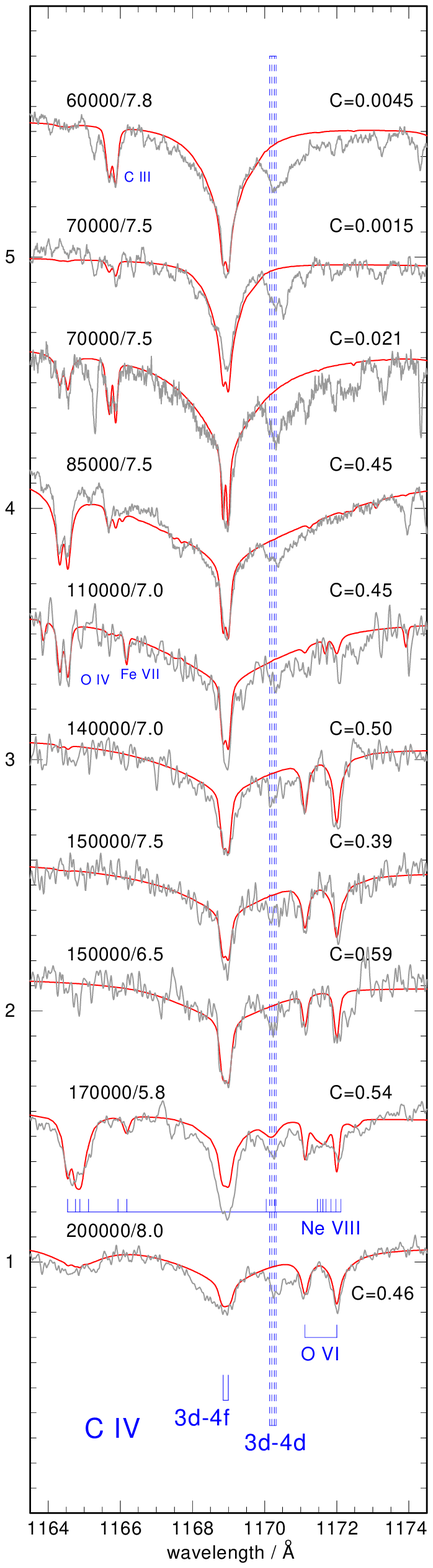}
  \caption{Observed spectra of DO white dwarfs and PG\,1159 stars
    (grey lines) with overplotted model spectra (red lines; forbidden
    components not included). \emph{Left panel:} Forbidden
    \ion{C}{4} 3p--4f multiplet (indicated by vertical, blue dashed
    lines) in the blue wing of the allowed 3p--4d line. Object names
    and spectral types are indicated. \emph{Right panel:} For the same
    stars, we show the forbidden \ion{C}{4} 3d--4d multiplet in the
    red wing of the allowed 3d--4f line. \Teff, \logg, and carbon abundance (mass fraction)
    of the models are indicated. Some other photospheric lines are
    marked in both panels.\label{fig:all}  }
\end{figure*}

The most prominent example for numerous forbidden components are
neutral helium lines in optical spectra of white dwarfs
\citep[e.g.,][]{1976ApJ...204L.119L,1995ApJ...441L..85B}, originally
detected in B-type stars by \citet{1929ApJ....69..173S} at
\ion{He}{1}~$\lambda$4471\,\AA. This $2\,^3P-4\,^3D$ transition is
accompanied by the forbidden $\Delta \ell = 2$ component
$2\,^3P-4\,^3F$. Detailed descriptions of the physical process
associated with the formation of forbidden components in neutral
helium can be found, e.g., in
\citet{1969A&A.....1...28B,1974JQSRT..14.1025B,1974slbp.book.....G,1991JQSRT..45...11A,2005pps..book.....G}.

To our best knowledge, forbidden components of elements heavier than
helium have hitherto not been identified in astrophysical plasmas
but in
laboratory plasmas. \citet{1987PhRvA..36.2265B} report the
detection of such components in lithium-like (i.e., one valence
electron) \ion{C}{4} and \ion{N}{5}. They can be used for plasma
diagnostics because they are strongly density dependent. Another
example is lithium itself. \ion{Li}{1} $\lambda$4602\,\AA\ 2p--4d with its
forbidden 2p--4p and 2p--4f components is used for diagnostics
\citep[e.g.,][]{2014AcSpe.100...86C} in plasmas where helium with
its forbidden components is not present for that purpose.

The first to investigate the presence of forbidden line components in
stellar spectra was, as mentioned, \citet{1929ApJ....69..173S}. He
wrote: ``Of the various elements only helium seems to promise any
results. Hydrogen shows no new lines outside the Balmer components,
which are blended. All other elements are either faint in the stars or
not very susceptible to Stark effect.''

At last we can report here on the detection of forbidden line
components of a heavier species, namely carbon in stellar
spectra. They occur in hot (pre-) white dwarfs, namely, the same two
\ion{C}{4} transitions discovered in the plasma experiment by
\citet{1987PhRvA..36.2265B}. The detection is favored by the
conditions encountered in hot white dwarf atmospheres. Broad
\ion{C}{4} lines due to strong Stark effect and, often, highly
enriched carbon abundance. Because of the strong density dependence of
the line strengths, they can potentially be used as sensitive gravity
indicators in stellar spectra.

\begin{figure}[bth]
 \centering\includegraphics[width=0.95\columnwidth]{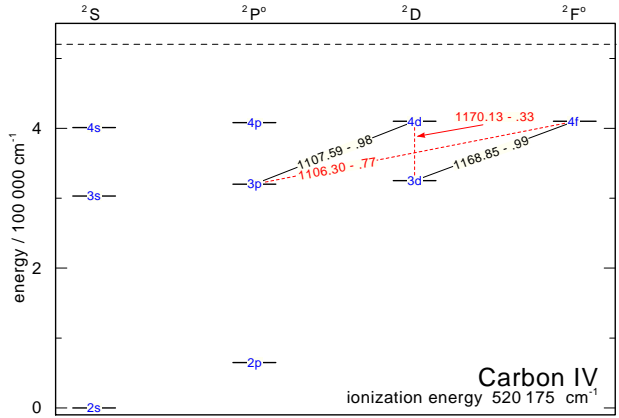}
  \caption{Grotrian diagram of the lithium-like \ion{C}{4} ion. Solid
    lines indicate the observed dipole allowed transitions, dashed
    lines indicate the identified forbidden
    components.}\label{fig:grotrian}
\end{figure}

\section{Detection of forbidden \ion{C}{4} components}

We have detected two forbidden \ion{C}{4} components in ultraviolet
spectra of three DO (i.e., He-dominated) white dwarfs and seven
PG\,1159 (He-C-O dominated) stars (Fig.\,\ref{fig:all}). The DOs'
effective temperature and surface gravity ranges are \Teff =
$60\,000-70\,000$\,K and $\log (g/{\rm cm/s}^2)=7.5-7.8$, and their carbon abundances
range between C = 0.0015 and 0.021 (mass fraction). The PG\,1159 stars
are hotter (\Teff = $85\,000-200\,000$\,K) and have surface gravities
between \logg = 6.5 and 8. Their carbon abundances are significantly
higher (C = $0.39-0.59$). All stars have in common that the profiles of
the allowed \ion{C}{4} transitions are very broad because of linear Stark effect
in comparison to most other metal lines (e.g., from \ion{C}{3} and
\ion{O}{4} visible in some spectra displayed in
Fig.\,\ref{fig:all}).

The detected forbidden components are 3p--4f ($\Delta \ell = 2$;
Fig.\,\ref{fig:grotrian}) in the blue wing of the allowed 3p--4d line
at 1108\,\AA, and 3d--4d ($\Delta \ell = 0$) in the red wing of the
allowed 3d--4f line at 1169\,\AA. Their profiles are asymmetric, with
the broader wing pointing away from the adjacent allowed line. In some
spectra, the 3p--4f transition can be resolved as two fine-structure
components. The spacing between the allowed and forbidden components
is about 0.5\,\AA. All \ion{C}{4} lines are multiplets with noticeable
splittings because the energy levels are doublets. In
Table~\ref{tab:lines}, we present the wavelengths of the forbidden
components as calculated from the level energies listed in the
National Institute of Standards and Technology (NIST)
database\footnote{\url{http://www.nist.gov/pml/data/asd.cfm}} together
with new pseudo-relativistic Hartree-Fock oscillator strengths
computed using the Cowan's atomic structure codes \citep{cowan1981}.

We are confident that the identification of the two absorption
features as forbidden \ion{C}{4} components  is correct and that they
do not stem from other elements. First, none of the ions hitherto
identified in the investigated stars is visible over the entire, large
\Teff\ range covered by the stars. Second, the observed asymmetric
line profiles are neither expected nor observed from allowed transitions.

We noticed the forbidden lines during previous analyses of some of the
stars presented here, however, as yet they remained unidentified. They
are most prominent in the DO white dwarfs (top three spectra in
Fig.\,\ref{fig:all}) and generally weaker in the hottest PG\,1159
stars. As can be judged from the model for RX\,J2117.1+3412, the
3d--4d line in this star is blended by a \ion{Ne}{8} line, however,
the other forbidden component 3p--4f is clearly visible.

The displayed spectra are overplotted with models whose relevant
parameters (\Teff, \logg, carbon abundance) are given in the right
panel of Fig.\,\ref{fig:all}. Most of them were derived in our earlier
work, whilst a few are from work in progress, involving new
observations (see below). For conciseness we abstain from individual
references. As representative examples we mention our detailed work on
the DO white dwarf \re\ \citep[][and references
  therein]{2016A&A...590A.128R} and on two PG\,1159 stars
\citep{2015A&A...582A..94W}. As to the observations, the majority of
the spectra from the objects in the present study were recorded with
the \emph{Far Ultraviolet Spectroscopic Explorer} (FUSE). All were
described in the publications mentioned. Spectra of PG\,0111+002,
PG\,0109+111, and PG\,1707+427, however, are from our observations
recently performed with \emph{Cosmic Origins Spectrograph} aboard the
\emph{Hubble Space Telescope} (Proposal ID 13769). Details of these
observations and their spectral analysis will be deferred to a later
paper (Hoyer et al., in prep.).

\begin{table}
\begin{center}
\caption{Wavelengths $\lambda$ and oscillator strengths $f$ of the two
  forbidden \ion{C}{4} components' multiplets.}
\label{tab:lines} 
\begin{tabular}{crcc}
\hline 
\hline 
\noalign{\smallskip}
$n\ell-n'\ell'$ & $j-j'$   & $f$                  & $\lambda$/\AA \\
\noalign{\smallskip}
\hline 
\noalign{\smallskip}
3p--4f          & 1/2--5/2 & $1.199\times10^{-5}$ & 1106.40 \\
                & 3/2--5/2 & $1.730\times10^{-6}$ & 1106.79 \\
                &    --7/2 & $1.042\times10^{-5}$ & 1106.77 \\
\noalign{\smallskip}
3d--4d          & 3/2--3/2 & $6.886\times10^{-7}$ & 1170.20 \\
                &    --5/2 & $2.937\times10^{-7}$ & 1170.14 \\
                & 5/2--3/2 & $1.958\times10^{-7}$ & 1170.32 \\ 
                &    --5/2 & $7.977\times10^{-7}$ & 1170.27 \\ 
\hline
\end{tabular} 
\end{center}
\end{table}

\begin{figure*}[bth]
 \centering\includegraphics[height=0.33\textwidth]{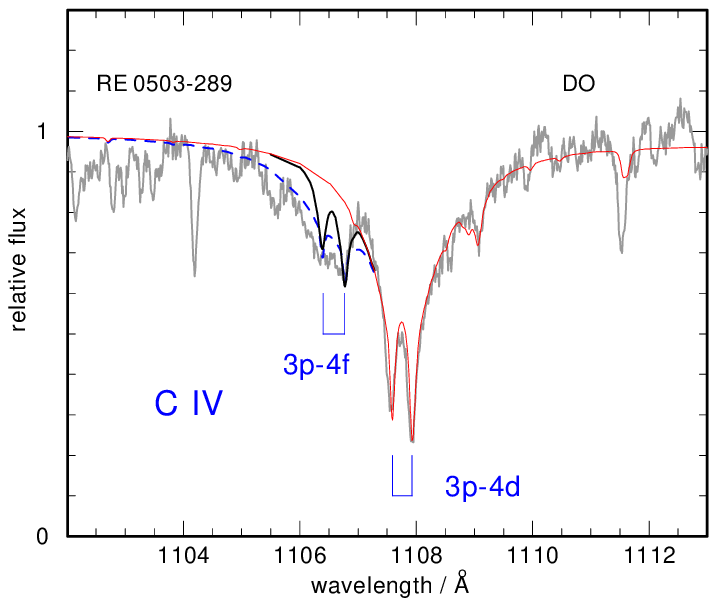}\hspace{0.5cm}
           \includegraphics[height=0.33\textwidth]{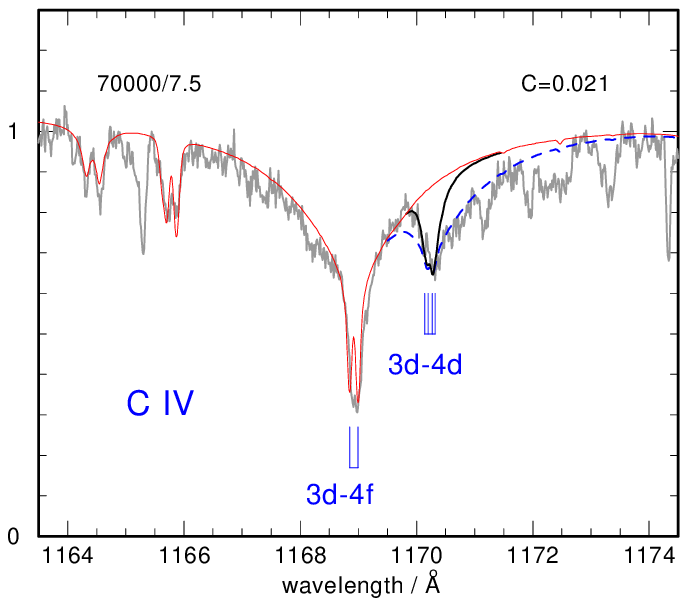}
  \caption{Details from Fig.\,\ref{fig:all} showing the DO white dwarf
    \re\ and a model without the forbidden components (thin, red
    line). In addition, two more models that include the \ion{C}{4}
    forbidden components in an approximate manner (see text) are
    overplotted. Thick, black line: $f$-values
    artificially increased. Thick, blue dashed line: Stark broadening
    parameter increased.}\label{fig:all_det}
\end{figure*}

\section{Discussion and conclusions}

Appropriate quantum-mechanical calculations for the Stark line
broadening of forbidden \ion{C}{4} components are not
available. \citet{1994PhRvE..49.5889G} performed such calculations for
the interpretation of laboratory spectra with a computer code
presented by \citet{1990PhRvA..42.5433C} but, to our best knowledge,
they were not published. For the 3p--4f transition, broadening data
were published by \citet{1991A&AS...89..581D}, however, they are
useless in our context because it was assumed an isolated line. At the
moment, we are only able to include the lines by assuming linear Stark
effect (as for the allowed components) in the approximation presented
by \citet{1991A&A...244..437W} and guessing their strength by
arbitrarily upscaling theoretical $f$-values computed by us. As an
example we show in Fig.\,\ref{fig:all_det} (black line) the result of
this procedure in the case of the DO white dwarf \re. The $f$-values
required scaling by factors of 400 and 10\,000 for the 3p--4f and
3d--4d transitions, such that they amount to about $f =
0.004-0.008$. The line positions are matched while the asymmetries are
not because, obviously, our assumption for broadening is poor. The
extended wings that point away from the allowed line component are not
broad enough in the model. An arbitrary increase of the Stark damping
constant by a factor of six and a further increase of the $f$-values
by a factor of two results in a better fit, but then the steep wings
pointing towards the allowed line components are too broad
(Fig.\,\ref{fig:all_det}, dashed blue line).

The asymmetric line shape of the forbidden components (with their
steep wing always towards the adjacent allowed line) is identical to
the behaviour of such lines of neutral helium in stellar
atmospheres. Under certain circumstances, the asymmetry can be very
pronounced as was demonstrated by \citet[][their
  Fig.\,2]{1998ApJ...496..395B} and was explained by the effect of
varying ratios of line widths to the separation of forbidden and
allowed components as a function of formation depths of line cores and
wings.

The detection of the \ion{C}{4} 3d--4d forbidden component at
1171\,\AA\ in the experiment by \citet{1987PhRvA..36.2265B} (observed
in second order) was subsequently questioned by
\citet{1994PhRvE..49.5644G} who argued that the respective spectral
feature is an impurity line, namely the $\lambda$585\,\AA\ \ion{C}{3}
2p\,3s--2p$^2$ line (at 1171\,\AA\ in fourth order). In light of our
observations in white dwarfs we conclude that
\citet{1987PhRvA..36.2265B} indeed saw the \ion{C}{4} 3d--4d
component, at least contributing to the impurity line.

The forbidden components of \ion{C}{4} originally identified in
laboratory plasmas \citep{1987PhRvA..36.2265B} and now in stellar
spectra are $n = 3-4$ transitions, where $n$ is the principal quantum
number. The 3d--4d forbidden component of lithium-like \ion{N}{5} was
also detected by \citet{1987PhRvA..36.2265B}. It is located at
749\,\AA\ and therefore not accessible in stellar spectra because of
extinction by interstellar neutral hydrogen. The respective $n = 3-4$
lines of lithium-like \ion{O}{6} have even shorter
wavelengths. \ion{O}{6} has strong and broad lines, comparable to the
\ion{C}{4} lines, in the ultraviolet spectra of the PG\,1159 stars
presented here. Candidates for forbidden components are $n = 4-5$
transitions, but we could not identify any.

\acknowledgments

DH and TR are supported by the German Aerospace Center (DLR) under
grants 50\,OR\,1501 and 50\,OR\,1507. Financial support from the
Belgian FRS-FNRS is also acknowledged.  PQ is research director of
this organization. The TMAD service
(\url{http://astro-uni-tuebingen.de/~TMAD}) used to compile atomic
data for this paper was constructed as part of the activities of the
German Astrophysical Virtual Observatory. This research has made use
of the SIMBAD database, operated at CDS, Strasbourg, France, and of
NASA's Astrophysics Data System Bibliographic Services.  Some/all of
the data presented in this paper were obtained from the Mikulski
Archive for Space Telescopes (MAST). 


\vspace{5mm}
\facilities{HST(STIS,COS), FUSE}

\bibliographystyle{astroads}

\end{document}